\documentclass[aps,prl,twocolumn,superscriptaddress,showpacs,amsmath,amssymb,longbibliography]{revtex4-2}

\usepackage{xcolor}
\usepackage{graphicx}
\usepackage{dcolumn}
\usepackage{bm}
\usepackage{amsmath}
\usepackage{breqn}
\usepackage{subfigure}
\usepackage{float}
\usepackage{braket}
\usepackage{times}
\usepackage{comment}
\usepackage[colorlinks,linkcolor=blue,citecolor=blue,urlcolor=blue]{hyperref}
\newcommand{\ESM}{\mbox{E}_{\mbox{\scriptsize SM}}}
\newcommand{\EFM}{\mbox{E}_{\mbox{\scriptsize FM}}}
\newcommand{\EDMI}{\mbox{E}_{\mbox{\scriptsize DMI}}}
\usepackage{amsmath}
\usepackage[bottom]{footmisc}

\DeclareMathOperator{\arccot}{arccot}

\begin{document}

\preprint{APS/123-QED}

\title{Magnon Superlattices around Skyrmions in Frustrated Magnets}

\author{Adarsh Hullahalli$^{1,2}$, Christos Panagopoulos$^{1}$, and Christina Psaroudaki$^{2}$
\\
\vspace{3mm}
\textit{$^{1}$Division of Physics and Applied Physics, School of Physical and Mathematical Sciences, Nanyang Technological University, 637371 Singapore} \\
\textit{$^{2}$Laboratoire de Physique de l'École Normale Supérieure, ENS, Université PSL, CNRS, Sorbonne Université, Université de Paris, F-75005 Paris, France}}

\date{\today}

\begin{abstract}
Dynamic and stable magnetic textures offer a powerful platform for controlling magnon states in the broader context of spin electronics. In this work, we uncover a novel class of dynamical, crystal-like localization patterns in real space, arising from the hybridization of magnons with topologically non-trivial spin textures that possess helicity as an internal degree of freedom. By tuning the magnon wavelength to match the size of these textures, specifically, atomic-scale skyrmions in centrosymmetric frustrated magnets, we achieve strong interference effects. This leads to the emergence of magnon superlattices, shaped by the internal skyrmion structure and the underlying Mexican-hat magnon dispersion. Furthermore, helicity-driven nonlinear dynamics give rise to dispersive magnon bands with nontrivial Chern numbers within the first magnon gap. These findings provide fundamental insights into magnon behavior in complex spin environments and establish frustrated magnets as a versatile platform for manipulating spin excitations at the atomic scale.
\end{abstract}

\maketitle

\section{\label{sec:level1} Introduction}

\begin{figure*}[t]
    \centering
    \includegraphics[width=1\linewidth]{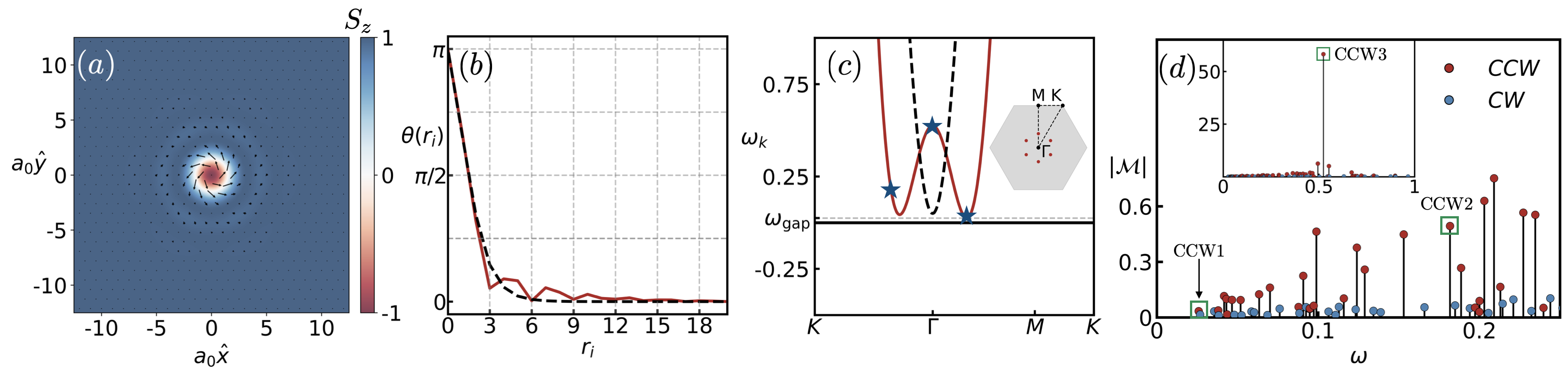}
     \caption{\textbf{Finite-wavelength magnon excitations matching skyrmion size.} (a) Isolated skyrmion of a frustrated magnet in a triangular lattice with $J_1 = 1$, $J_2 = 0.5$, $h = 0.225/S$, and $K = 0.15$. (b) Skyrmion profile calculated along a line passing through the skyrmion center as $\theta(r_i) = \arccos{S^z_i}$ for isolated skyrmions in frustrated (red line) and chiral magnets (black line), which illustrates the skyrmion tails, unique to frustrated magnets. (c) Low energy dispersion of magnon excitations around the uniformly magnetized state of a frustrated (red line) and chiral magnet (black line) along the $\Gamma$-M-K direction. The Mexican hat-like shape in frustrated magnets introduces complexity to the magnon dynamics around skyrmions. (d) Magnetic dipole moment $|\mathcal{M}|$ of the magnon modes around a single skyrmion.}
     \label{Isolated_Skyrmion}
 \end{figure*}

The interaction of multiple length scales can give rise to hybridization effects and emergent phenomena that transcend the properties of individual constituents. When characteristic lengths, such as domain sizes, coherence lengths, or moiré pattern periodicities, become comparable, superlattice structures with fundamentally new functionalities emerge. This convergence of scales enables the design and control of complex quantum states, including unconventional superconductivity and topologically nontrivial phases, opening pathways toward materials with tailored electronic, magnetic, and optical properties\cite{Dagotto2005,RevModPhys.86.1189,Tokura2018,Cao2018,Balents2020}. 

Competing magnetic interactions and length scales can give rise to new states of matter. Magnetic skyrmions, topologically protected spin textures, have emerged as a key example due to their potential applications in spintronics and quantum computing \cite{Fert2017,2410.11427}. Although skyrmions are commonly found in non-centrosymmetric materials with strong Dzyaloshinskii-Moriya interactions (DMI), they can also form in centrosymmetric frustrated magnets due to competing exchange interactions \cite{Leonov2015}. Unlike DMI-based skyrmions, which arise from broken inversion symmetry and exhibit fixed chirality, frustrated magnets allow for smaller and chirality-flexible structures \cite{PhysRevB.93.064430}. This inherent ground-state degeneracy makes them a unique platform for exploring exotic spin patterns and quantum effects\cite{PhysRevLett.127.067201,PhysRevB.106.104422}. 

Understanding the magnon spectrum—the collective spin-wave excitations—around skyrmions is essential for unraveling their fundamental physics and assessing their potential for information storage and processing \cite{Garst_2017}. These excitations play a key role in skyrmion stability and provide a means to manipulate skyrmion motion and interactions through external stimuli \cite{PhysRevB.90.094423,Mochizuki2014,PhysRevB.104.054419}. The magnon spectrum around a skyrmion can exhibit distinct features such as chiral edge modes \cite{PhysRevResearch.2.013231} and finely spaced emergent Landau levels \cite{doi:10.1126/science.abe4441}. The interplay between magnons and skyrmions can lead to emergent phenomena, including non-reciprocal behavior and long-range propagation characteristics \cite{Seki2020}. Understanding these features has driven exciting advances in experimental techniques for probing spin-wave excitations, such as inelastic neutron scattering \cite{Portnichenko2016}, Brillouin light scattering \cite{2404.14314}, and ferromagnetic resonance \cite{PhysRevB.95.224405,Satywali2021}.

Collective spin excitations in frustrated magnets differ from those in DMI-based systems due to the unique properties of their skyrmions. The additional helicity degree of freedom couples to magnon modes, enabling rich hybridization dynamics. Because skyrmions in frustrated magnets are typically smaller, the number of internal deformation modes they can support is limited. The skyrmion core induces spatially modulated fluctuations in its vicinity that mediate long-range interactions and reshape the magnon propagation landscape.

We begin by studying the magnon spectrum around an isolated skyrmion and a skyrmion lattice. The small size of frustrated skyrmions confines localized deformation modes to the magnon continuum, where they hybridize with extended states. A novel magnon superlattice pattern arises from interference effects, becoming observable through the unique coupling between the skyrmion’s magnetically active modes and magnon modes with a characteristic Mexican-hat dispersion. 
This reveals a novel paradigm of emergent confinement in magnonic systems, a highly tunable dynamical mechanism that provides valuable insights into how topological textures shape quasiparticle interference in quantum materials.  

While the system hosts several counterclockwise (CCW) modes, it lacks magnetically active breathing modes due to its $S_z$-conserving $U(1)$ symmetry.  Because the Hamiltonian is invariant under global rotations around the $z$ axis, the longitudinal spin is conserved within linear order. The breathing modes have a vanishing magnetic dipole moment and are inactive in linear response, but can be weakly excited through nonlinear helicity-mediated coupling.

Yet, these modes are excited through second-order coupling mechanisms beyond the linear response regime. Magnon excitations in frustrated magnets hybridize with the helicity-zero mode, particularly in breathing modes, leading to helicity precession. Bands in skyrmion lattices exhibit complex $k$-dependence and are characterized by finite Chern numbers within the first magnon gap, with a sign that depends on skyrmion charge. In systems of interacting skyrmions, emergent low-energy modes are characterized by complex dynamics related to the precession of skyrmion helicity. 

\begin{figure*}[t]
    \centering
    \includegraphics[width=1\linewidth]{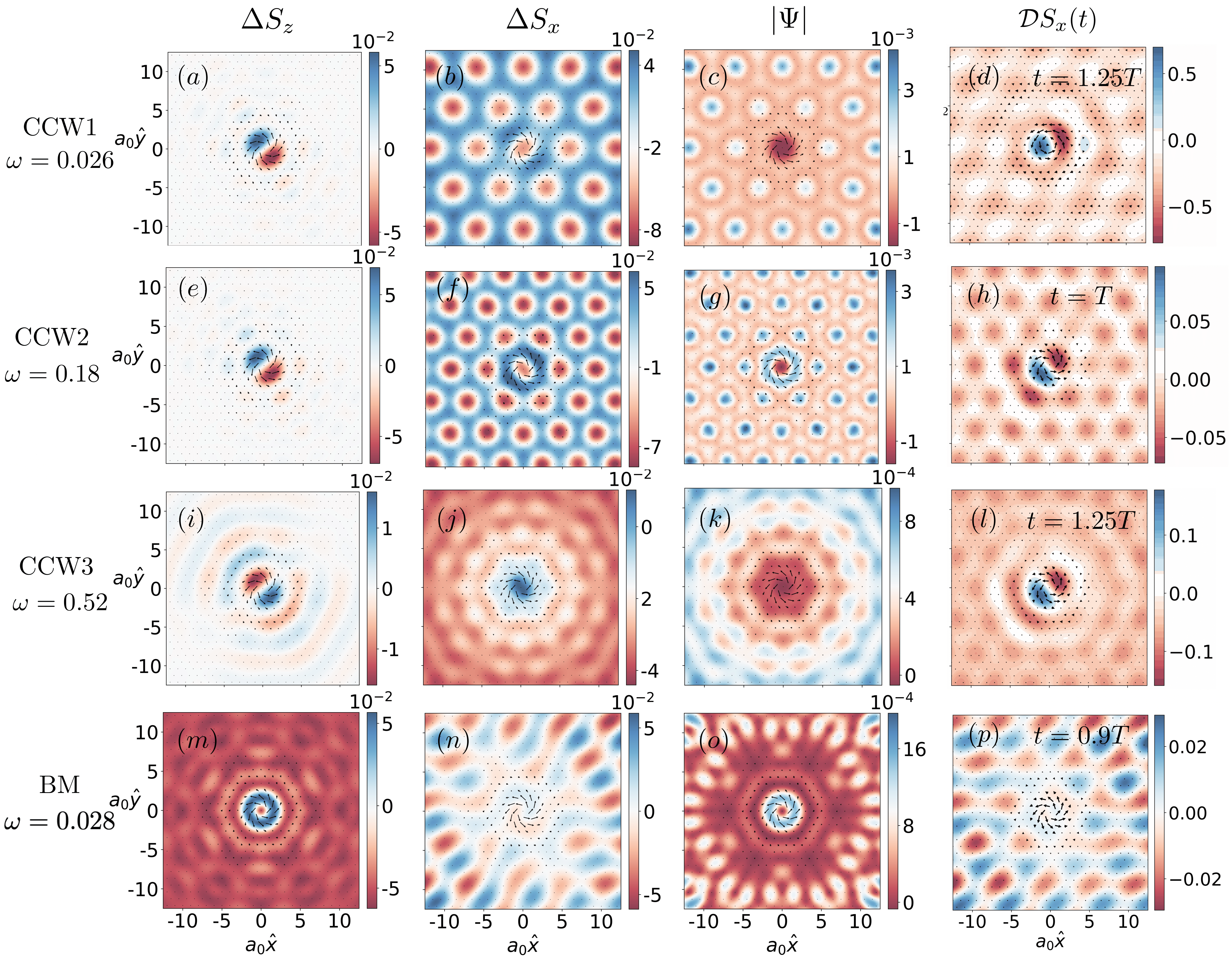}
     \caption{\textbf{Observable magnon superlattices.} Crystal-like localization patterns of magnon modes in the presence of a skyrmion. (a)-(d) Real-space deformation $\Delta S_z$ and $\Delta S_x$, probability density $\vert \Psi \vert$, and real-time dynamics $\mathcal{D} S_x(t)$ for the low-lying $\mbox{CCW1}$ hybridized with interference patterns of finite-wavelength extended states. (e)-(h) Same quantities for the $\mbox{CCW2}$ mode, exhibiting magnon crystals with shorter periodicity. (i)-(k) The symmetric localization patterns of the high in energy $\mbox{CCW3}$ mode with the largest $\mathcal{M}_x$. (m)-(p) The lowest lying breathing mode has a vanishing $\mathcal{M}_z$, but is dynamically activated due to a second-order effect. Here we use the parameters of Fig.~\ref{Isolated_Skyrmion}.}
     \label{Localization}
 \end{figure*}
 
\section*{\label{sec:Model}Results}

\textbf{Model}. We consider classical spins, $\mathbf{S}_i$, defined on a two-dimensional (2D) $x-y$ lattice with ferromagnetic (FM) nearest-neighbour (NN) and competing skyrmion-stabilizing spin interactions $\ESM$,
\begin{equation}
    H = -J_1 \sum_{\langle i, j \rangle} \mathbf{S_i} \cdot \mathbf{S_j} +\ESM - h \sum_i S^z_i - K \sum_i (S^z_i)^2 \,, \label{eq:Hamiltonian}
\end{equation}
 where $\langle i, j \rangle$ sums over nearest-neighbor pairs, while the third and fourth terms represent the interaction with the magnetic field aligned along the $z$-axis and the single-ion magnetic anisotropy, respectively. Here we focus on the magnetization dynamics in the presence of competing antiferromagnetic interactions on a triangular lattice of the form $\ESM = \EFM=
  J_2 \sum_{\langle \langle i, j \rangle \rangle} \mathbf{S_i} \cdot \mathbf{S_j}$, where $\langle \langle i, j \rangle \rangle$ denote pairs of next nearest-neighbor pairs. When necessary, we compute the magnon spectrum in chiral skyrmions with DMI $\ESM=\EDMI=\sum_{\langle i, j \rangle} \mathbf{D}_{ij} \cdot \mathbf{S_i} \times \mathbf{S_j} $, to facilitate comparison and highlight the distinct nature of magnons in frustrated magnets.

The zero-temperature phase diagram of the model \eqref{eq:Hamiltonian} as a function of $h$ and $K$ includes eight distinct phases, comprising a skyrmion crystal and a fully polarized FM state \cite{Leonov2015,PhysRevB.93.184413}. In the continuum limit,  $\textbf{m}_i=\mathbf{S}_i/S$, with $S$ the total spin,  becomes a field $\mathbf{m}(r)$, typically  expressed in spherical coordinates as $\mathbf{m}=[\sin \Theta \cos \Phi,\sin \Theta \sin \Phi, \cos\Theta]$. Isolated skyrmions appear as metastable defects above the FM state, described by $\Phi=\mu \phi +\varphi_0$ and $\Theta(\mathbf{r})=\Theta(\rho)$, in polar coordinates $\mathbf{r}=(\rho,\phi)$, where $\varphi_0$ is the helicity and $\mu$ the vorticity. These topological solutions have integer charge $Q=\int d\mathbf{r}~\mathbf{m} \cdot (\partial_x \mathbf{m} \times \partial_y \mathbf{m})/(4\pi)=-\mu$ assuming a FM background $\mathbf{m}_{\mbox{\scriptsize FM}}=(0,0,1)$. Since the energy is independent of $\varphi_0$ and the sign of $\mu$, skyrmions ($\mu=1$, $Q=-1$) and antiskyrmions ($\mu=-1$, $Q=1$) have the same energy and the helicity is a zero mode. 

\begin{figure*}[t]
    \centering
    \includegraphics[width=1\linewidth]{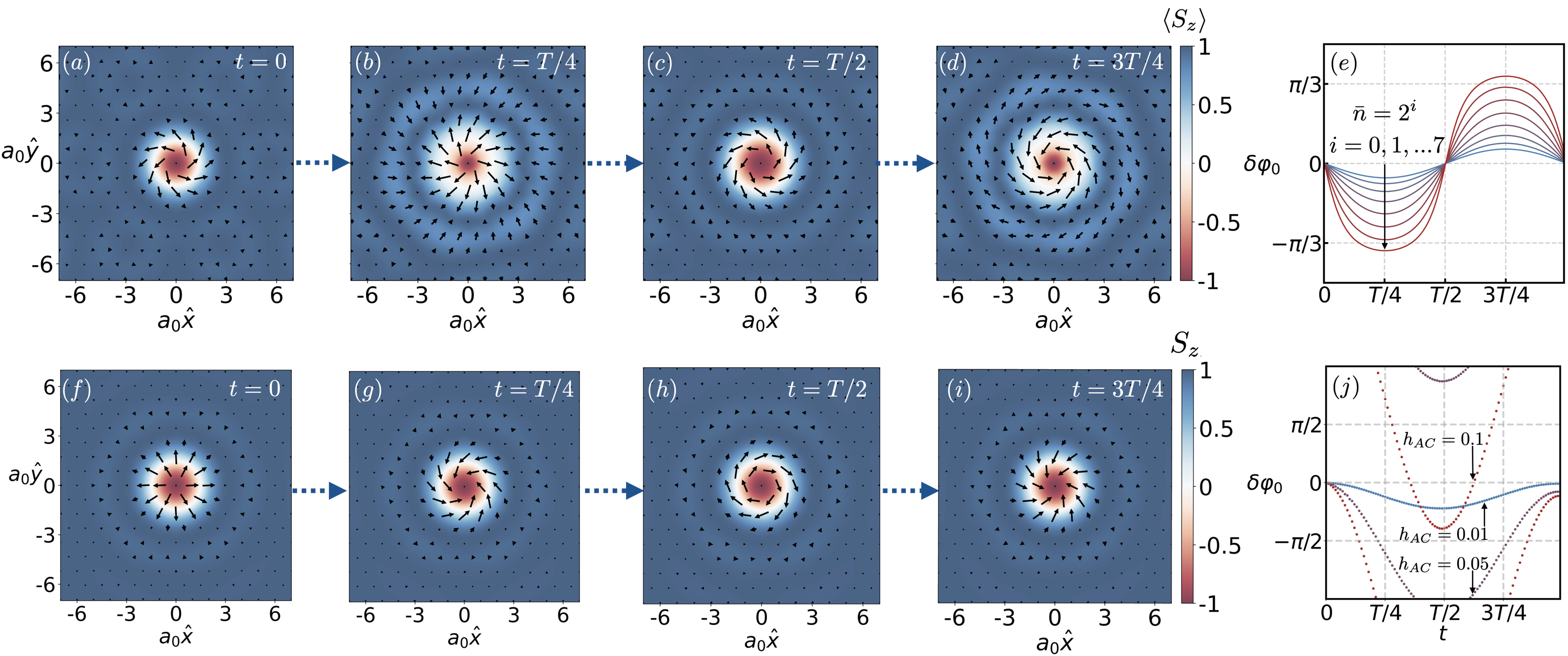}
     \caption{\textbf{Helicity induced nonlinear dynamics}. (a)-(d) Isolated skyrmion breathing mode from \ref{Localization} (m)-(p) involving coupled oscillation of skyrmion $z-$magnetization and helicity, $\varphi_0$, found at $\omega=0.028$. The average magnon number per lattice site is $\bar{n}/N^2 = 2^7/42^2 \approx 0.073$. Snapshots are taken at equally spaced time-steps and cover one full oscillation period. (e) Nonlinear dependence of helicity oscillations over one period on the magnon number $\bar{n}$. (f)-(i) Snapshots of the same breathing mode activated by an out-of-plane magnetic field $h_{\mbox{\scriptsize AC}} = 0.1$ due to second-order coupling. (j) Nonlinear dependence of helicity oscillations over one period on the drive amplitude.}
     \label{Breathing}
 \end{figure*}

\textbf{Methods}. The energy in Eq.~\eqref{eq:Hamiltonian} is minimized using an iterative simulated annealing procedure combined with single-step Monte Carlo dynamics based on the Metropolis algorithm \cite{Metropolis01091949}. Magnon modes are calculated around a stable classical state using linear spin-wave theory by applying the Holstein-Primakoff transformation \cite{PhysRev.58.1098} and an expansion in powers of $1/S$. The details of calculating and diagonalizing the spin wave (SW) Hamiltonian $H_{\mbox{\scriptsize SW}}$ \cite{COLPA1978327,Toth_2015} are reviewed in Appendix A\ref{sec:Magnons}. Simulations were conducted with periodic boundary conditions on triangular lattices of size $42\times42$, unless stated otherwise. The magnetization dynamics are obtained by numerically solving the Landau-Lifshitz-Gilbert (LLG) equation  \cite{TATARA2008213} using the fourth-order Runge–Kutta method and a Gilbert damping parameter $\alpha=0.01$.

For clarity, we briefly summarize the quantities used throughout the work. The dynamic magnetic dipole moment $\mathcal{M}$ characterizes the coupling strength of each magnon mode to an oscillating magnetic field and identifies magnetically active excitations. The spatial deformation of a mode is visualized by 
$\Delta \mathbf{S}_{\mathbf{n}}$, which represents the real-space spin deviation in the coherent magnon representation relative to the static skyrmion background. The quantity $\mathcal{D}S(t)$ denotes the time-dependent deviation of the spin field obtained from micromagnetic simulations under an AC drive, while $\phi_0$ refers to the helicity angle of a skyrmion—the global rotation of in-plane spins around the $z$ axis and $\varphi$ defines the relative helicity between two skyrmions.
~\\
\textbf{Magnon Superlattices}. We first consider the case of an isolated skyrmion, a metastable state within the uniformly magnetized phase for fields above the critical value $h_c=J_2-2K$ with its spin configuration shown in Fig. \ref{Isolated_Skyrmion} (a). The skyrmion profile $\theta_i=\arccos S^z_i$ depicted in Fig.~\ref{Isolated_Skyrmion} (b) (red solid line) exhibits oscillations and decays away from the skyrmion core \cite{PhysRevB.93.064430}. Therefore, the skyrmion tail can be regarded as an excitation in the FM state, induced by the nonlinear core and extending into the topologically trivial region surrounding the skyrmion.  For the parameters considered, the skyrmion core diameter is $\lambda=12a$, with $a$ the lattice constant, determined from the points where $\theta_i$ vanishes, and it depends only weakly on the applied magnetic field. This oscillatory behavior is absent in DMI skyrmions, with an exponentially decaying skyrmion profile shown in Fig.~\ref{Isolated_Skyrmion} (b) (black dashed line). 

Small-amplitude spin waves in the FM state appear as traveling waves with a dispersion $\omega_{\mathbf{k}}$ depicted in Fig.~\ref{Isolated_Skyrmion} (c). The energy-momentum cut along the high-symmetry $\Gamma$-M-K direction exhibits a Mexican hat-like shape \cite{PhysRevX.7.041049,PhysRevB.101.134420}. Along this path, the low-energy magnon dispersion features one gapped minima, $k^{1}_{\mbox{\scriptsize min}}$ marked by red dots in the hexagonal Brillouin zone, with a gap given by $\omega_{\mbox{\scriptsize gap}}=h+2K-J_2$. In total, there are six degenerate minima, corresponding to spin waves with wavevectors matching the characteristic length scale of the model (\ref{eq:Hamiltonian}) and propagating along the lattice vectors of the 2D triangular lattice. In contrast, spin waves in DMI systems propagate with a quadratic dispersion $\omega_k \propto k^2$ above the magnon gap, shown in Fig.~\ref{Isolated_Skyrmion} (c) with a black dashed line.  The plotted curve shows the symmetric branch $\omega_{|k|}$, thus the dispersion shift at a finite $k_D$ caused by the DMI term \cite{KATAOKA1983341} is not visible as the opposite $\pm k$ contributions overlap.

Similar to other topological objects \cite{PhysRevB.52.7412,PhysRevB.89.024415}, the presence of a skyrmion creates a confining potential for magnons, giving rise to localized modes that correspond to deformations. In chiral magnets, the wavelength of the low-energy magnon continuum is considerably larger than the skyrmion size ($l_{\mbox{\scriptsize min}} =2\pi/k_{\mbox{\scriptsize min}}\gg \lambda$), such that the skyrmion barely perturbs the extended magnon modes \cite{PhysRevB.89.024415}. In frustrated magnets, however, the situation is qualitatively different. Here, the low-energy magnons have a finite wavelength comparable to the skyrmion size, $l_{\mbox{\scriptsize min}} \sim \lambda$, enabling strong coupling between the two. Since the skyrmion size is small, it supports no internal excitations below $\omega_{\mbox{\scriptsize gap}}$. We find several deformation-related states that lie within the magnon continuum and are hybridized with extended modes. 

We now analyze the magnon modes, concentrating on those that are magnetically active, characterized by a finite magnetic dipole moment $\boldsymbol{\mathcal{M}}_{\mu}$ \cite{PRXQuantum.3.040321} with $\mu=\{x,y,z\}$. In chiral skyrmions, magnetic resonance studies investigating the response to a uniform oscillatory magnetic field have identified three characteristic in-gap magnon modes: a clockwise (CW) and CCW gyration of the skyrmion core, associated with a finite $\mathcal{M}_{x,y}$ induced by in-plane AC fields, and a breathing mode with a finite $\mathcal{M}_{z}$ which emerges under longitudinal AC fields \cite{PhysRevLett.108.017601,Garst_2017,PhysRevLett.109.037603}. Antiskyrmion resonances are found at the same frequencies as skyrmion modes, but with opposite sense of gyration (CCW skyrmion modes correspond to CW antiskyrmion modes and vice versa) \cite{10.1021/acs.nanolett.0c02192}. We therefore focus on skyrmions with $Q=-1$ and omit the discussion of antiskyrmions with $Q=1$.

Above the magnon gap, we identify the CCW, CW, breathing, and multipolar modes as hybridized excitations of the skyrmion and the surrounding FM region. As a result, these modes are influenced by the distinctive dispersion of the uniformly magnetized state [see Fig. \ref{Isolated_Skyrmion} (c)]. The lowest mode near $\omega_{\mbox{\scriptsize gap}}$ is a CCW mode with a finite $\mathcal{M}_x$ [CCW1 mode in  Fig.~\ref{Isolated_Skyrmion} (d)]. Fig.~\ref{Localization} illustrates the real-space deformation $\Delta S_z$ (a) and (b) $\Delta S_x$ defined as $\Delta \mathbf{S}_{\mathbf{n},i} = \langle \mathbf{S}_{\mathbf{n},i}\rangle-S\mathbf{v}^3_i$, with $\mathbf{v}^3_i$ a vector aligned with the classical skyrmion state $\mathbf{S}_{i,0}$. $\langle \mathbf{S}_{\mathbf{n},i}\rangle$ is over the coherent magnon representation \cite{PhysRevResearch.2.013231,PhysRevB.4.201} (see Appendix B for definitions). Notably, $\Delta S_x$ exhibits a crystal-like structure away from the skyrmion core, which arises due to interference effects between the six degenerate extended states with $k^{i}_{\mbox{\scriptsize min}}$ with $i=\{1,6\}$ that exist at the same energy [see Fig.~\ref{Isolated_Skyrmion} (c)]. The crystal-like structures correspond to localization patterns of the probability density of the wavefunction $\vert \Psi_{\mbox{\scriptsize CCW1}} \vert ^2$, shown in Fig.~\ref{Localization} (c). These interference patterns emerge from the wave-like behavior of magnons, further confirmed by the equal superposition of magnon states $\vert \Psi \rangle= \sum_{i=1,6} \vert k^{i}_{\min} \rangle$ plotted in Fig.~\ref{superposition} (a) in the Appendix in the coherent state representation, revealing the emergence of a crystal-like phase with the same periodicity. Thus, the coherent mixing of extended states with different momenta manifests as spatially modulated localization superlattices, with an amplitude that survives far away from the skyrmion core [Fig~.\ref{superposition} (b)]. Such emergent localization is a hallmark of strong magnon-skyrmion coupling in frustrated systems and has no analog in conventional chiral magnets, where the mismatch in length scales suppresses this interplay.


\begin{figure}[t]
    \centering
    \includegraphics[width=1\linewidth]{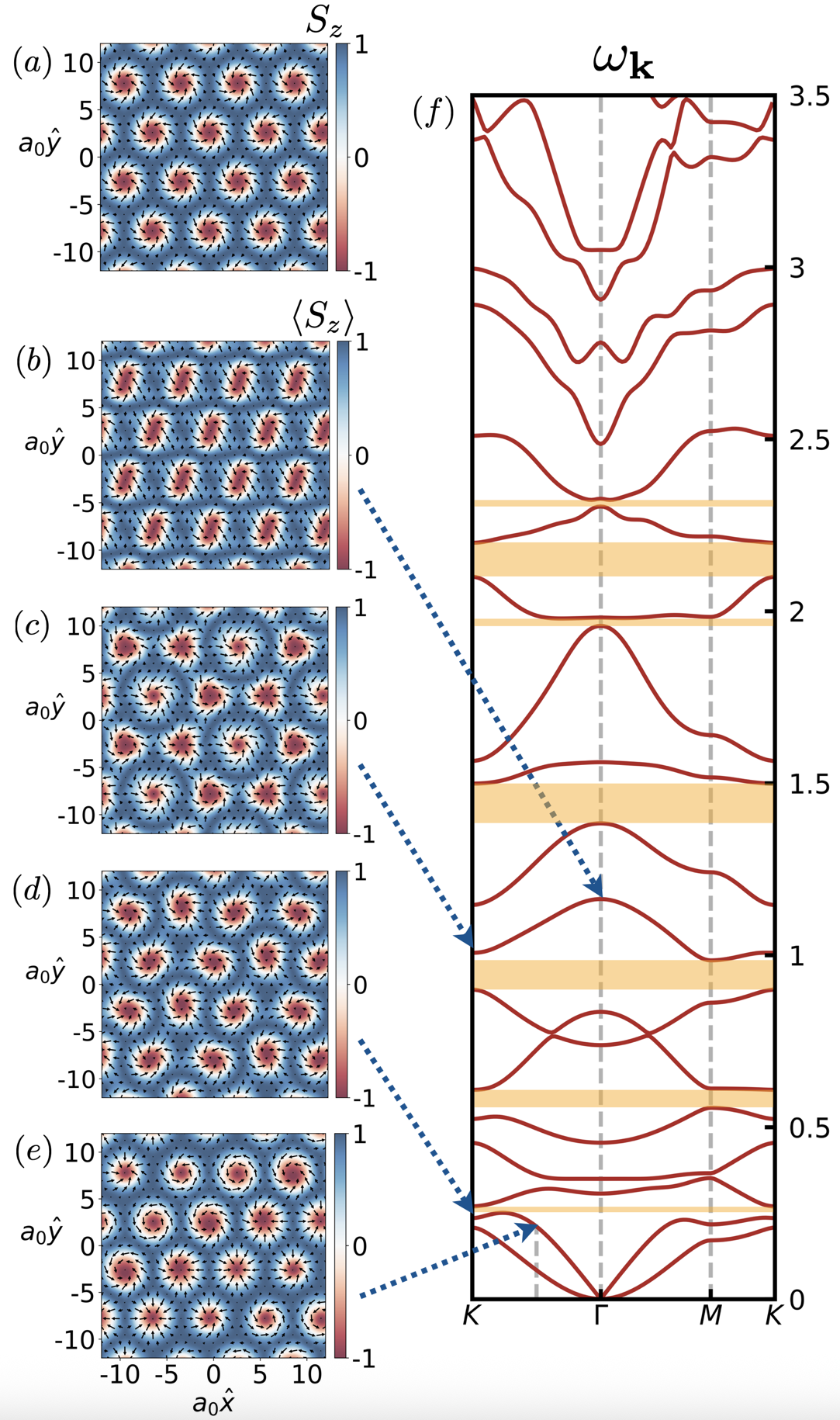}
    \caption{\textbf{Magnon bands around skyrmion lattice.}(a) Skyrmion lattice ground state at parameters: $J_1=1$, $J_2=0.5$, $h = 0.15/S$, $K=0.15$. (b)-(e) Snapshots of excited skyrmions calculated with $\bar{n}_{\mu, \mathbf{k}} = 3$ magnons per unit cell. The symmetry of multipolar modes is dependent on magnon momentum (b, c). A new band corresponding to helicity excitations emerges, and hybridizes with multipolar channels at large $\mathbf{k}$ (d, e). (f) Magnon band structure of the skyrmion lattice. Band gaps are highlighted in gold, with edges corresponding to frequencies of topologically-protected edge states (see Figure 5). Due to complex skyrmion-skyrmion interactions, multipolar bands are also dispersive in frustrated magnets.}
    \label{Lattice_Bands}
\end{figure}

\begin{figure*}[t]
    \centering
    \includegraphics[width=0.95\linewidth]{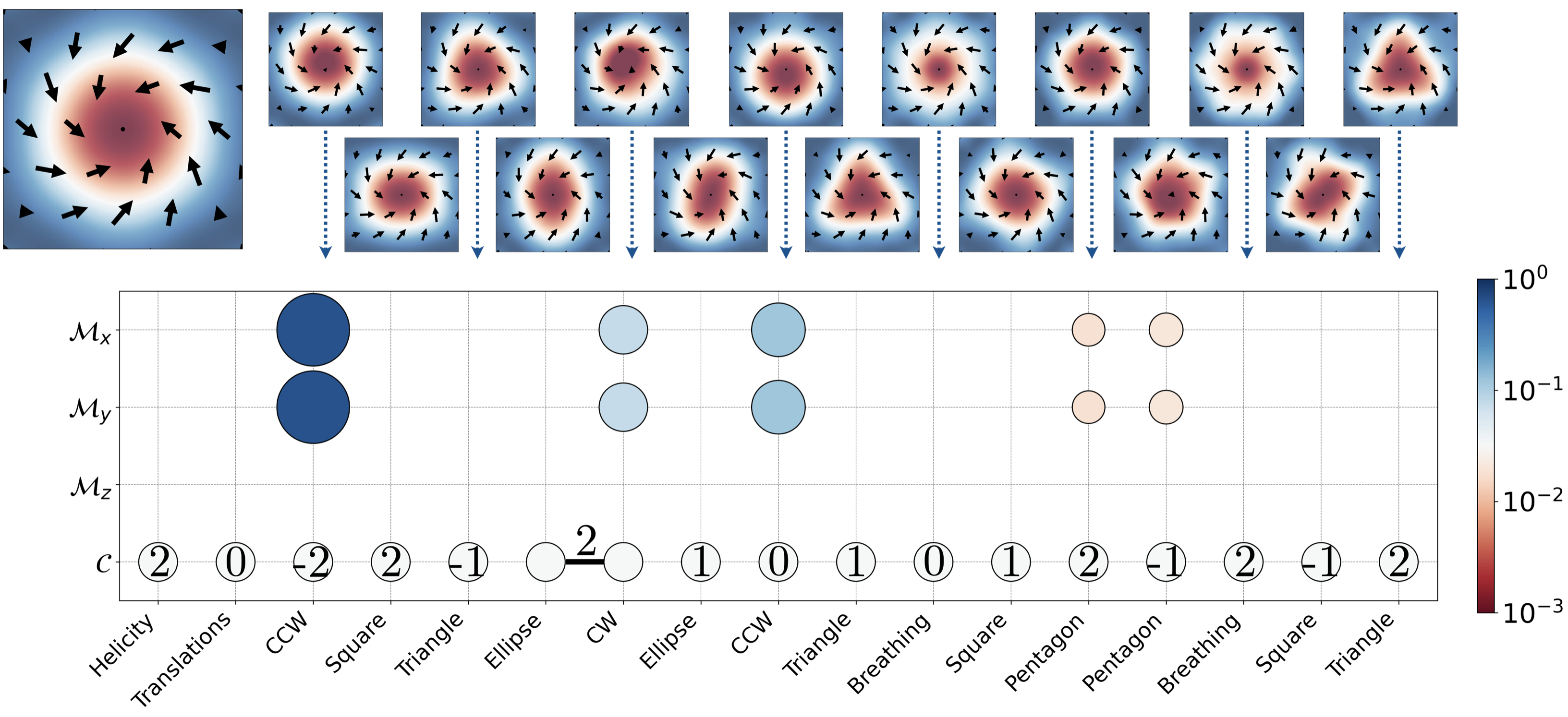}
    \caption{\textbf{Topological magnon bands}. Magnetic dipole moment $\boldsymbol{\mathcal{M}}_\mu$ and topological Chern numbers $\mathcal{C}$ of the lowest magnon bands shown in Fig.~\ref{Lattice_Bands}. For antiskyrmion lattices, the sign of $\mathcal{C}$ flips, illustrating the dependence of edge state chirality on bulk topology. The neighboring elliptical and CCW bands touch, and have a total Chern number of $2$. Some topological edge states are observable via in-plane AC magnetic field excitation due to their finite magnetic dipole moment.}
    \label{Chern_numbers}
\end{figure*}

To investigate whether this novel magnon mode localization leads to observable patterns, we numerically solve the LLG equation \cite{TATARA2008213} for magnetization dynamics under $\mathbf{H}_{\mbox{\scriptsize AC}}=h_{\mbox{\scriptsize AC}}\cos(\omega t) \hat{\mathbf{x}}$ with frequency $\omega=\omega_{\mbox{\scriptsize gap}}$, and a single skyrmion as an initial state. Fig.~\ref{Localization} (d) presents snapshots of the time evolution of magnetization deviation $\mathcal{D}S_x(t)=S_x(t=0)-S_x(t)$ using $h_{\mbox{\scriptsize AC}}=0.1$, showing that the AC field not only excites the CCW gyration of the skyrmion core but also induces crystal-like long-range localization patterns. The full-time evolution is shown in Supplementary Movie 1. This localization phenomenon becomes \textit{observable} from the unique interplay between the skyrmion’s magnetically active modes and the characteristic Mexican-hat magnon dispersion. Notably, the crystal patterns are entirely absent when starting from an FM state or in DMI skyrmion systems (see Fig.~\ref{CCW_DMI} in the Appendix for a comparison). 

The patterns are richer at higher energies since they contain both short- and long-wavelength components. Fig.~\ref{Localization} (e)-(f) shows the higher-energy CCW2 mode, which forms magnon crystals with shorter periodicities due to interference of shorter-wavelength magnons. The higher in-energy CCW3 mode has the largest magnetic moment $\mathcal{M}_x$ and is hybridized with modes at the $\Gamma$ point and symmetric localization patterns [Fig.~\ref{Localization} (i),(j)]. Thus, the interference conditions modify the patterns and lead to regimes of tunable, dynamic magnon localization. We confirm that these structures are observable via LLG simulations [Fig.~\ref{Localization} (h) and Supplementary Movie 2 using $h_{\mbox{\scriptsize AC}}=0.01$ and $\omega=0.18$ for the CCW2 mode and (l) and Supplementary Movie 3 using $h_{\mbox{\scriptsize AC}}=0.001$ and $\omega=0.52$ for CCW3]. In contrast to DMI skyrmions, there is no breathing mode with finite $\mathcal{M}_z$, due to the $S_z$-conserving $U(1)$ symmetry of the model. Focusing on the lowest breathing mode with $\omega_{\mbox{\scriptsize BM}} \approx \omega_{\mbox{\scriptsize gap}}$, as shown in Fig.~\ref{Localization} (m)-(n), we conclude that the out-of-plane dipole moment of the skyrmion core $\sum_{i < \lambda} \mathcal{M}^{i}_z$ cancels out by the opposite dipole moment of the tail, $\sum_{i> \lambda} \mathcal{M}^{i}_z$, shown in Fig~\ref{dipole_BM}. 

To verify that the interference mechanism is not an artifact of perfect rotational symmetry, we performed additional simulations including rotational-symmetry-breaking terms such as dipole–dipole interactions and weak DMI exchange. As shown in Fig.\ref{Symmetry_Break} of Appendix D, the characteristic real-space modulations remain intact. The magnon superlattices persist with only minor amplitude distortions and unchanged periodicity, confirming that these patterns originate from the finite-$k_{\min}$ structure of the frustrated-magnet dispersion and the strong magnon–skyrmion hybridaztion.

~\\
\textbf{\normalsize{Experimental feasibility and detection requirements}}. The dynamical magnon superlattices predicted here manifest as spatially modulated interference patterns whose periodicity is set by the characteristic magnon wavelength $l_{\min}$, comparable to the skyrmion diameter $\lambda$. Prime material platforms are centrosymmetric frustrated magnets in which $\lambda\sim l_{\mbox{\scriptsize min}}$. Notable examples include Gd$_2$PdSi$_3$ \cite{science.aau0968}, Gd$_3$Ru$_4$Al$_{12}$ \cite{Hirschberger2019}, and GdRu$_2$Si$_2$ \cite{Khanh2020}, which host nanometric skyrmion lattices with periods in the few nm range, as well as kagome magnets such as Fe$_3$Sn$_2$ \cite{Hou2017SkyrmionBubbles}, where skyrmionic textures are stabilized at room temperature. Their small skyrmion size ensures strong magnon–skyrmion hybridization, while the availability of bulk single crystals and thin films makes them compatible with high-resolution techniques. In realistic units of $J_1=2$ meV, $J_2=1.5$ meV, $K=0.3$ meV, $B=h J_1/(g \mu_B)=3.45$ T, and $a=0.5$ nm, the magnon gap energy appears at $\omega_{\text{gap}}=0.05$ meV [12 GHz], while the skyrmion core is $\lambda=6$ nm. Resolving these patterns requires imaging techniques with spatial resolution of a few nanometers, achievable with spin-polarized scanning tunneling microscopy (STM) \cite{PhysRevLett.88.057201,Wiesendanger2009SPSTM}, magnetic exchange force microscopy (MExFM) \cite{Wiesendanger2009SPSTM}, scanning transmission X-ray microscopy \cite{Mamyrbayev2019}, or nitrogen-vacancy (NV) center magnetometry with a demonstrated NV-to-sample distance around 8 nm \cite{Xu2025NVdistance}. 

In reciprocal space, the CCW modes occur at frequencies $5$–$30\%$ above the magnon gap (10-20 GHz), corresponding to absolute separations of roughly 1-10 GHz for realistic exchange parameters. Although in-plane AC fields can selectively excite CCW modes, their frequencies are close enough that resolving them individually, and thus identifying the corresponding interference patterns, requires sub-GHz spectroscopic resolution. Such performance is well within the reach of Brillouin light scattering \cite{fphy.2015.00035}, time-resolved magneto-optical Kerr effect \cite{PhysRevLett.96.217202}, and X-ray ferromagnetic resonance \cite{PhysRevLett.117.087205}. Importantly, the predicted interference patterns are robust and persist far from the skyrmion core [Fig.~\ref{superposition}(b)], allowing detection in both real-space and momentum-resolved measurements. 

A transverse spin deviation $\mathcal{D}S_x/S=0.3$ produces a transverse field at the surface $\Delta B_{\perp}=\mu_0 M_s \mathcal{D}S_x/S=0.38$ T, where we used $M_s=10^6$ A/m. The field decays exponentially with distance from the surface, giving estimated amplitudes of 40 mT at 10 nm from the surface. Such periodic stray-field modulations are detectable with spin-polarized STM and MExFM techniques which can directly image the in-plane spin contrast at the surface. Systematic spectroscopic and imaging studies in these materials can directly access the interference-driven localization patterns predicted here. 

Experimental \cite{science.aau0968} and first-principles studies \cite{PhysRevLett.125.117204} of Gd-based skyrmion hosting materials demonstrated that $J_{1,2} \gg J_{\perp}$, with $J_{\perp}$ the interlayer coupling, leading to magnetically quasi-two-dimensional behavior. A weak $J_{\perp}$ does not modify the in-plane finite-$k$ structure or the interference mechanism that gives rise to the magnon superlattices.

~\\
\textbf{\normalsize{Collective helicity precession modes}}. In frustrated magnets, magnon excitations hybridize with the helicity zero mode, with the strongest effects observed for the breathing modes. In Fig.~\ref{Breathing} (a)-(d), we depict snapshots of the time evolution of the out-of-plane magnetization $\langle S_{\mathbf{n},i}^z \rangle$ associated with the breathing mode at $\omega=0.028$. In Fig.~\ref{Breathing} (e), we plot dynamical deviations of the skyrmion helicity $\delta \varphi_0 (t)= \varphi_0(t) - \varphi_0(t=0)$, defined as
\begin{align}
\varphi_0(t)=\arg \left(\sum_{i}\sqrt{1-S_i^z(t)}\sqrt{1-S_{i,0}^z} e^{i (\Phi_i(t)-\Phi_{i,0})}\right)\,,
\end{align}
where $\Phi_i=\arccot(S_i^x/S_i^y)$, and $S^z_{i,0}$,$\Phi_{i,0}$ the spin configuration of the initial static skyrmion. The extent of helicity precession depends on the magnon number and saturates at sufficiently large $\bar{n}_i$. In Fig.~\ref{Breathing} (f)-(i) we show snapshots of $\mathcal{D}S_i^z(t)$ obtained by solving the LLG equation under an AC drive of amplitude $h_{\mbox{\scriptsize AC}}=0.1$ and $\omega=0.028$. We observe a similar oscillation of the skyrmion helicity Fig.~\ref{Breathing} (j), with an amplitude that increases with $h_{\mbox{\scriptsize AC}}$. 

The vanishing $\mathcal{M}_z$ for all breathing modes implies that an out-of-plane AC field cannot directly activate them. Nevertheless, numerical simulations of the LLG equation under an AC field $\mathbf{H}_{\mbox{\scriptsize AC}}=h_{\mbox{\scriptsize AC}}\cos(\omega_{\mbox{\scriptsize BM}} t) \hat{\mathbf{z}}$ reveal that the breathing mode is weakly excited [Fig~\ref{Localization}-(p)] through a second-order coupling mechanism beyond the linear response regime. Here, the AC field couples to the total skyrmion spin $\mathcal{S}^z_0= \sum_{i}(S-S_{i,0}^z)$ and activates the oscillation of the skyrmion helicity $\varphi_0(t)$, depicted in Fig.~\ref{Breathing} (j). Due to the strong hybridization between breathing modes and helicity [see Fig.\ref{Breathing} (a)-(d)], this time-dependent helicity parametrically modulates the magnon mode profiles, leading to their weak excitation shown in Fig.~\ref{Localization} (p). This is a second-order effect beyond linear order, resulting from mode hybridization.

Taken together, these results establish the basis for understanding magnon behavior in the vicinity of a single skyrmion. 
The following sections on topological magnon bands and coupled-skyrmion dynamics complete this picture by illustrating how these behaviors evolve from isolated textures to periodic or interacting assemblies.

~\\
\textbf{\normalsize{Topological Magnon Bands}}. The periodicity of the 2D hexagonal skyrmion crystal gives rise to a magnon band structure with a corresponding hexagonal Brillouin zone. Figure \ref{Lattice_Bands} shows the lowest-energy magnon bands along the $\Gamma$-M-K high-symmetry path. All bands are dispersive due to strong interactions between neighboring skyrmions, particularly the overlap of their tails. Since the skyrmion crystal breaks both translational symmetry and global spin rotational symmetry along the magnetic field axis, two Goldstone modes are expected at the $\Gamma$ point, where the two lowest bands touch zero energy. Due to the discreteness of the spin lattice, a small finite gap appears at $\mathbf{k} = 0$. In addition, a new band emerges above the lowest translational mode, corresponding to precessional dynamics of the skyrmion helicity around its equilibrium value [see Fig.~\ref{Lattice_Bands} (b)].

Within the dispersive multipolar bands, we observe symmetry changes across the Brillouin zone. For example, as shown in Fig.~\ref{Lattice_Bands} (b), a mode exhibiting twofold rotational symmetry at the $\Gamma$ point transitions to a threefold rotational symmetry at the K point. Similarly, at the edges of the Brillouin zone, the helicity band corresponds to elliptical skyrmion deformations with no change in helicity. Notably, chiral skyrmion lattices support several flat bands associated with non-dispersive, localized skyrmion distortions \cite{PhysRevResearch.2.013231}. These flat bands arise due to suppression of skyrmion-skyrmion interactions in the small-skyrmion limit. In contrast, frustrated skyrmions exhibit fluctuating tails that mediate long-range skyrmion-skyrmion interactions \cite{Leonov2015}, leading to a qualitatively different band structure.

We note that each band $m$ can be classified by the following Berry curvature
\begin{align}
\mathcal{B}_m(\textbf{k})=i\epsilon_{\mu \nu} \mbox{Tr}[P_m(\mathbf{k})[\partial_{k_\mu}P_m(\mathbf{k})]\partial_{k_\nu}P_m(\mathbf{k})]\,,
\end{align}
with $P_m=\mathcal{T}_k \Pi_m \sigma_z\mathcal{T}_k^{\dagger}\sigma_z$ a projection operator, $(\Pi_m)_{ij}=\delta _{mi}\delta_{ij}$, and $\mathcal{T}_k$ the paraunitary Bogoliubov transformation that diagonalizes the spin-wave Hamiltonian. The corresponding topological Chern index is defined as an integral over the first Brillouin zone $\mathcal{C}_m=\int d\mathbf{k} \mathcal{B}_m(\mathbf{k})/2\pi$ \cite{PhysRevB.87.174427}. Topological magnon bands characterized by finite Chern numbers and associated chiral edge modes have emerged as a central theme in magnonics \cite{PhysRevB.87.174427}. A particularly rich platform for such phenomena is the skyrmion lattice phase \cite{PhysRevLett.122.187203} which supports unidirectional magnon edge states robust against disorder \cite{PhysRevResearch.2.013231}. Nevertheless, intrinsic magnon interactions can compromise the topological protection, leading to the breakdown of chiral edge magnons \cite{PhysRevB.109.024441}. Here we find a number of bands with non-vanishing Chern number presented in Fig.~\ref{Chern_numbers}. In particular, we note that the low-lying band of helicity excitations is characterized by $\mathcal{C}=2$ and is accompanied by a global gap,  an aspect which is absent in chiral skyrmion lattices. Note that the sign of the Chern number is dictated by the skyrmion’s topological charge. In particular, we find that magnon bands in antiskyrmion lattices exhibit Chern numbers of opposite sign, indicating that the band topology can be engineered by tuning the underlying skyrmion topology.

\begin{figure*}[t]
    \centering
    \includegraphics[width=1\linewidth]{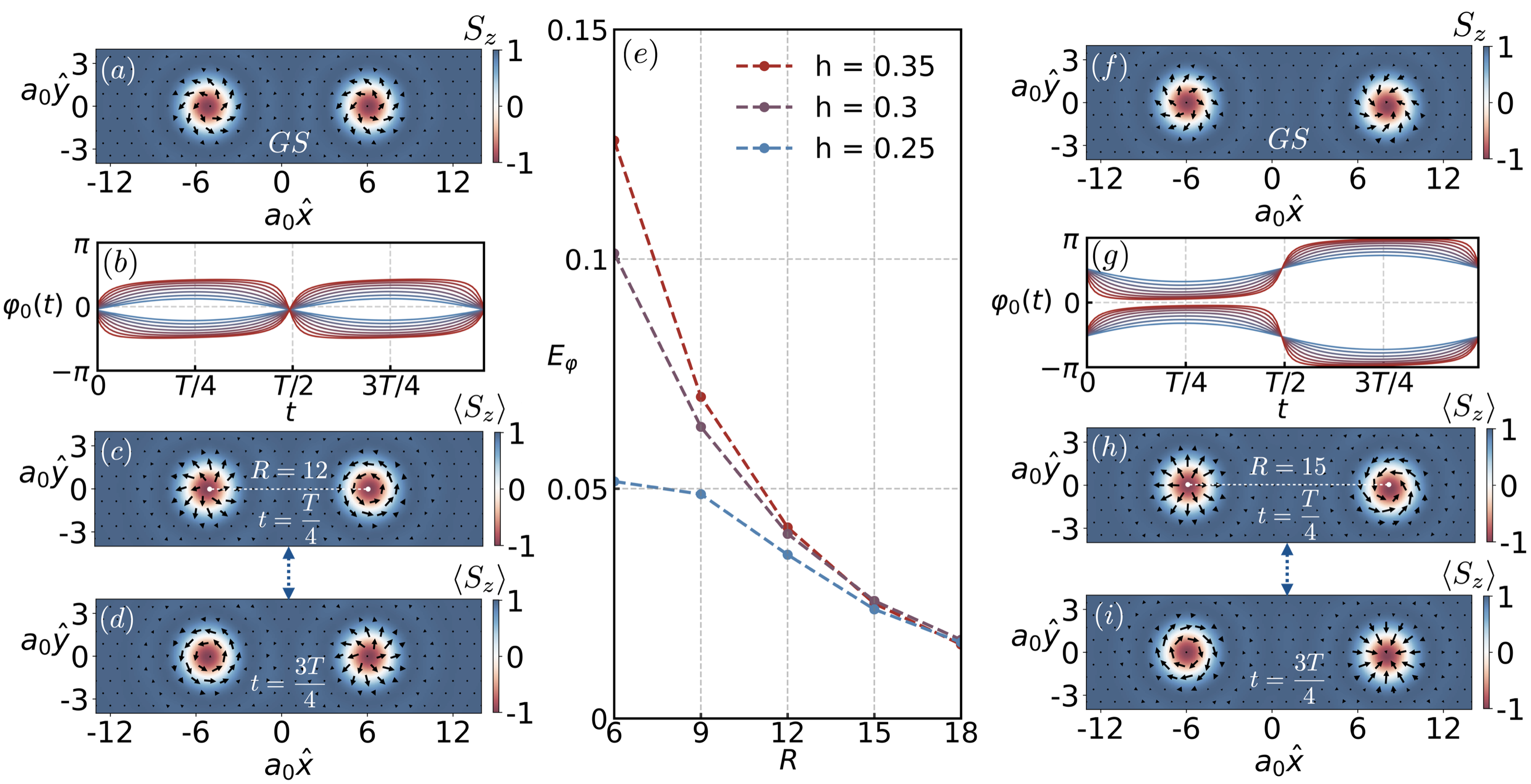}
    \caption{\textbf{Helicity excitation of coupled skyrmions}. The pairwise skyrmion interaction depends on the separation distance $R$ and the relative skyrmion helicity $\varphi$. (a) When $R=12$, the two skyrmions have the same helicity such that $\varphi=0$, while (b)-(d) the joint precession of skyrmion helicity occurs in opposite directions. (f) For $R=15$, the two skyrmions have the opposite helicities such that $\varphi=\pi$, and (g)-(i) helicity precession takes place along the same direction of rotation. (b) and (g) illustrate the dependence of $\varphi_0(t)$ over one period for different magnon numbers $\bar{n}$. Blue to red lines correspond to $\bar{n} = 2^i, i=0, 1,...7$.}
    \label{two_skyrmions}
\end{figure*}

~\\
\textbf{\normalsize{Coupled Dynamics of Interacting Skyrmions}}. Long-range interactions between skyrmions suggest the presence of coupled dynamics in systems of multiple skyrmions. To study this, we stabilized a system of two skyrmions, each with helicity $\varphi_i$ spaced at a distance $R=\vert \mathbf{R}\vert$, illustrated in Fig.~\ref{two_skyrmions} (a) and (f). The pairwise skyrmion interaction depends on separation and skyrmion helicity as $E_{\mbox{\scriptsize int}} \sim \mbox{Re}[e^{-q R}]\cos(\varphi_1-\varphi_2)$, with $q$ a complex number that depends on the model parameters \cite{PhysRevB.93.064430}, implying that the relative skyrmion helicity $\varphi=\varphi_1-\varphi_2$ varies with $\mathbf{R}$. For the parameters considered here, we numerically find $\varphi=0$ when $ R=3n$, with $n$ an even integer, and $\varphi=\pi$ with $n$ an odd integer. The helicity precession is shown in Fig.~\ref{two_skyrmions} (b)-(d) for $R=12$ and  (g)-(i) for $R=15$. In the former case, we observe a joint precession of skyrmion helicity in opposite directions, while in the latter the two skyrmions precess along the same direction. Notably, similar to Fig.~\ref{Breathing}, the helicity excitations depend on the total magnon number $\bar{n}$. Fig.~\ref{two_skyrmions} (e) depicts the dependence of the energy of helicity excitation in the coupled system $E_{\varphi}$ as a function of separation $R$ for three values of the external magnetic field $h$. As the distance $R$ grows, the skyrmion interaction diminishes and helicity becomes a zero mode.

\section{\label{sec:Conclusion}Conclusions}

We studied dynamic magnonic patterns through geometric and topological engineering in frustrated magnets hosting atomic-scale size magnetic skyrmions possessing helicity as an internal degree of freedom. Our calculations of the magnon spectrum around isolated skyrmions, skyrmion lattices, and interacting skyrmions revealed that the magnon's wavelength is comparable to the skyrmion's size. This is distinct from chiral systems where skyrmion-induced effects on extended modes are typically weak due to a large mismatch in length scales. These results indicate strong magnon-skyrmion coupling, leading to emergent dynamical localization phenomena. Specifically, interference between degenerate extended states and the skyrmion CCW mode gives rise to robust magnon superlattices that survive far from the skyrmion core. 

Furthermore, we demonstrate that helicity acts as an internal dynamical degree of freedom, giving rise to nonlinear behavior such as helicity precession and mode hybridization. Although the model is $S_z$-conserving, magnetically inactive breathing modes are excited through second-order processes, mediated by the total skyrmion spin and helicity. In skyrmion lattices, skyrmion interactions give rise to complex dispersive magnon bands with nontrivial Chern numbers within the first bulk gap, confirming the topological nature of the excitations. We also uncover helicity-sensitive coupling between skyrmions, which gives rise to collective dynamics dependent on inter-skyrmion separation.

Our results reveal a previously unexplored regime of magnon physics, where frustration, topology, and internal skyrmion degrees of freedom coordinate to produce confined and topologically non-trivial spin excitations. These findings deepen the fundamental understanding of magnon behavior in complex spin textures and establish frustrated magnets as a fertile platform for studying emergent quasiparticles and their interactions. High energy techniques, imaging, and reconstruction algorithms are now critical to studying and visualizing the dynamical, crystal-like localization patterns we report. A complementary approach will be required, including high-frequency magneto-optics, x-ray ferromagnetic resonance, time-resolved momentum microscopy, and inelastic neutron scattering.

\section{Acknowledgments}
C.Ps. is an \'{E}cole Normale Sup\'{e}rieure (ENS)-Mitsubishi Heavy Industries (MHI) Chair of Quantum Information supported by MHI. The work in Singapore was supported by the Singapore Ministry of Education (MOE) Academic Research Fund Tier 3 grant MOE-MOET32023-0003.

\section{Correspondence}
Correspondence and requests for materials should be addressed to C.Ps. (email: christina.psaroudaki@phys.ens.fr) or to C.Pa. (email: christos@ntu.edu.sg).

\section{Contributions}
C.Ps. and C.Pa. supervised the project. A.H. developed the numerical code and performed the simulations and data analysis. CPs. and C.Pa. analyzed and interpreted the results and wrote the manuscript with the help of A.H.

\section{Competing Interests}
The authors declare no competing interests.

\section{Additional Information}
The online version of this file contains supplementary animations of the magnon dynamics (see "Supplementary Information"). Data is available on request from the authors.

\bibliography{apssamp}
~\\

\clearpage
\newpage
\begin{widetext}
\section*{Appendix}\label{sec:Appendix}
\subsection*{A. Magnetic Excitations}\label{sec:Magnons}
Skyrmion static solutions $\mathbf{S}_{i,0}$ emerge for a range of parameters in the centrosymmetric Hamiltonian of Eq.~\ref{eq:Hamiltonian}. Here, we calculate $\mathbf{S}_{i,0}$ using a Monte Carlo simulated annealing algorithm followed by solving the Landau-Lifshitz-Gilbert (LLG) equation with strong damping, as explained in Section Methods. We define a local orthonormal frame at each lattice site $i$ in terms of rotating orthonormal basis vectors $\{\mathbf{v}^1_i, \mathbf{v}^2_i, \mathbf{v}^3_i\}$, or $\mathbf{S}_i = \mathbf{v}^1_i S^1_i + \mathbf{v}^2_i S^2_i + \mathbf{v}^3_i S^3_i$, where $\mathbf{v}^3_i$ is aligned with the classical ground state $\mathbf{S}_{i,0}$.

\begin{figure}
    \centering
    \includegraphics[width=1\linewidth]{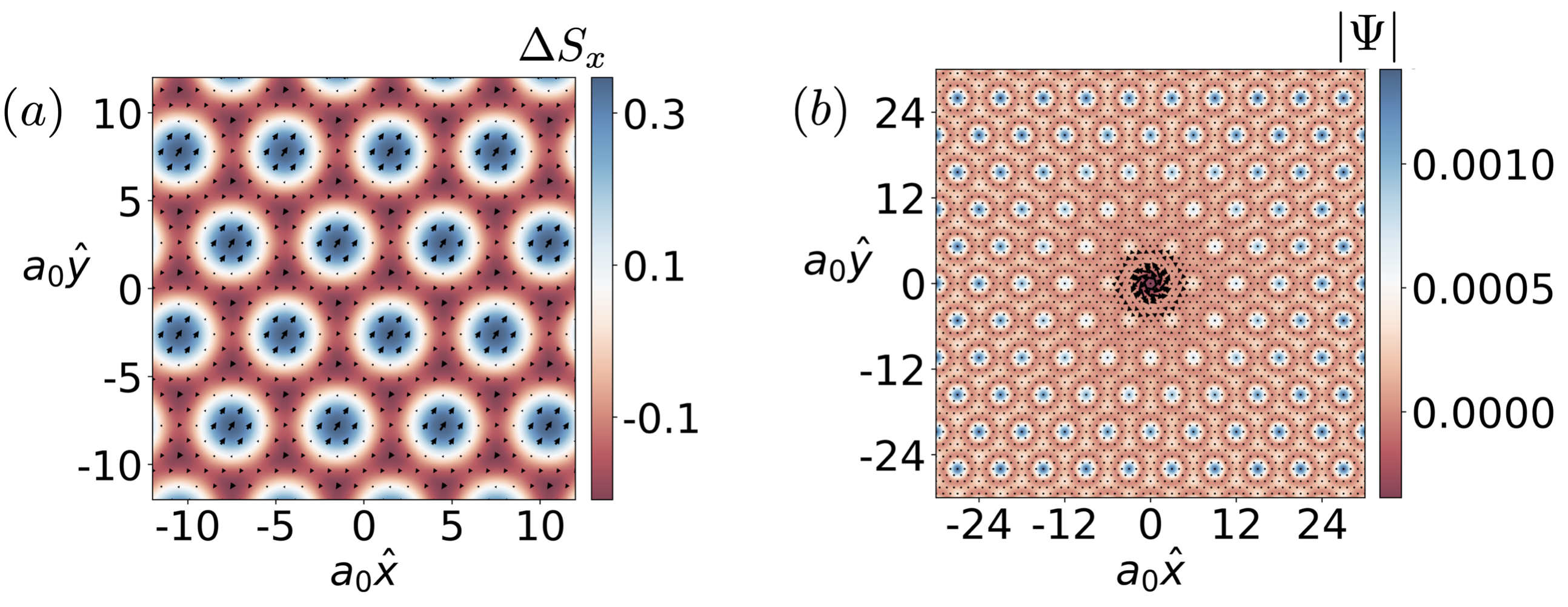}
    \caption{\textbf{Interference effects and superlattices.} (a) Real-space deformation $\Delta S_x$ of an equal superposition of the six low-lying degenerate extended states around the FM state of a frustrated magnet. A crystal-like phase emerges with the same periodicity as the dynamical magnon superlattices of Fig.~\ref{Localization}-(b). (b) Magnon amplitude $|\Psi|$ around an isolated skyrmion on a $72\times 72$ lattice with periodic boundary conditions, illustrating the survival of the superlattice pattern far away from the skyrmion core.}
    \label{superposition}
\end{figure}

Spin waves around $\mathbf{S}_{i,0}$ are estimated using an expansion of the form $\mathbf{S}_i/S=\sqrt{1-\vert \xi_i\vert^2}\mathbf{v}_3+\Re[\xi_i] \mathbf{v}_1+\Im[\xi_i] \mathbf{v}_2\approx\mathbf{S}_{i,0} + \delta \mathbf{S}_i$, with $\xi_i$ a complex field representing transverse fluctuations. Quantized magnetic fluctuations can be found by introducing magnon creation ($a^{\dagger}_i$) and annihilation ($a_i$) operators at each lattice site $i$ via the Holstein-Primakoff transformation, $S^3_i = S - a^{\dagger}_i a_i$, $S^+_i=S^1_i+iS^2_i = (2S - a^\dagger_ia_i)^{1/2}a_i$, and $S^-_i=S^1_i-iS^2_i = a^\dagger_i(2S - a^\dagger_ia_i)^{1/2}$. In this picture, linearized spin waves behave like harmonic oscillators described by bosonic operators $a_i\rightarrow \sqrt{S/2} \xi_i$ and $a_i^{\dagger}\rightarrow \sqrt{S/2} \xi_i^{*}$.  

For the case of a periodic structure with a magnetic unit cell of $N$ spins, the $2N \times 2N$ reciprocal-space spin wave Hamiltonian obtained from Eq.~\ref{eq:Hamiltonian} reads 
\begin{align}
    H_{\mbox{sw}} = \sum_{\mathbf{k}} \mathbf{A^\dagger_{\mathbf{k}} \mathcal{H}_{\mathbf{k}} A_{\mathbf{k}}}\,, \label{eq:MagnonH}
\end{align}
with $\mathbf{A_{\mathbf{k}}} =\left( a_{\mathbf{k}, i} ,a^\dagger_{\mathbf{-k}, i} \right)^{T}$
where reciprocal-space magnon operators are generated via $a^\dagger_{\mathbf{k}, i} =\sum_{\mathbf{n}} e^{i \mathbf{k} \cdot (\mathbf{r}_i + \mathbf{R_n})} {a}^\dagger_{\mathbf{n}, i}/{\sqrt{N}}$. Here, the position of a spin at sublattice index $i$ and unit cell index $\mathbf{n}$ is represented as $\mathbf{r}_i + \mathbf{R_n}$. The Hamiltonian has the structure:
\begin{align}
    \mathcal{H}_{\mathbf{k}} = \begin{pmatrix} \Phi^{-}_{\mathbf{k}} + C & \Phi^{+}_{\mathbf{k}} \\ (\Phi^{+})^*_{\mathbf{-k}} & (\Phi^{-})^*_{\mathbf{-k}} + C \end{pmatrix}\,, \label{eq:MagnonHk}
\end{align}
and components:

\begin{dmath}
    C_{\mu \nu} = -\delta_{\mu \nu} \left[J_1 \sum_{\langle (0,i), (\mathbf{n}, j) \rangle} \mathbf{v}^3_i \cdot \mathbf{v}^3_j + J_2 \sum_{\langle \langle (0,i), (\mathbf{n}, j) \rangle \rangle} \mathbf{v}^3_i \cdot \mathbf{v}^3_j + 2K (\mathbf{\hat{z}} \cdot \mathbf{v}^3_i)^2 + \frac{1}{S}h \mathbf{\hat{z}} \cdot \mathbf{v}^3_i\right]_{i=\mu} \label{eq:Cmn}
\end{dmath}

\begin{dmath}
    \Phi^{\pm}_{\mathbf{k}, \mu \nu} = \frac{1}{2} \left[J_1 \sum_{\mathbf{n}} \delta_{\mathbf{n}, j \in \langle0, i\rangle}e^{i \mathbf{k} \cdot (\mathbf{r}_j - \mathbf{r}_i + \mathbf{R_n})} \mathbf{v}^+_i \cdot \mathbf{v}^\pm_j + J_2 \sum_{\mathbf{n}} \delta_{\mathbf{n}, j \in \langle \langle0, i \rangle \rangle}e^{i \mathbf{k} \cdot (\mathbf{r}_j - \mathbf{r}_i + \mathbf{R_n})} \mathbf{v}^+_i \cdot \mathbf{v}^\pm_j + \delta_{ij} 2K (\mathbf{\hat{z}} \cdot \mathbf{v}^+_i) (\mathbf{\hat{z}} \cdot \mathbf{v}^\pm_j)\right]_{i=\mu, j=\nu}  \label{eq:Fkmn}
\end{dmath}

Note that equations \ref{eq:MagnonH},\ref{eq:MagnonHk} are general and recover the Hamiltonian for aperiodic systems in real space after setting $\mathbf{k}=0$. Diagonalizing Eq.~\ref{eq:MagnonH} \cite{COLPA1978327, Toth_2015} yields the Bogoliubov transformation matrix $\mathcal{T}^{\mathbf{k}}$ and energy matrix $E^{\mathbf{k}}$, which encodes information about $N$ unique eigenstates and satisfies $H^{\mathbf{k}}_{\mbox{sw}}\mathcal{T}^{\mathbf{k}}_{*, \mu} = E^{\mathbf{k}}_{\mu, \mu}\mathcal{T}^{\mathbf{k}}_{*, \mu}$, where $\mu = 0,1,2,...\ N - 1$ and $\mathcal{T}^{\mathbf{k}}_{*, \mu}$ is the column vector found at column $\mu$ of $\mathcal{T}$.
Matrix $\mathcal{T} =\begin{pmatrix} \boldsymbol{\psi}_u & \boldsymbol{\psi}_{v}^* \\ \boldsymbol{\psi}_{v} & \boldsymbol{\psi}_u^*\end{pmatrix}$, contains the magnon eigenvectors $\Psi^\mu = [\boldsymbol{\psi}^\mu_u, \boldsymbol{\psi}^\mu_v]^T$ in each column $\mu$. Resonance frequencies $\omega_{\mu}$ of modes $\Psi^\mu$ are elements $E^{\mathbf{k}}_{\mu, \mu}$ of the diagonal matrix $E$.

\subsection*{B. Magnon Coherent States}\label{sec:MagnonCoherent}
Following Ref.~\cite{PhysRevResearch.2.013231}, the real-space time evolution of the magnon modes is visualized using magnon coherent states,
\begin{align}
\vert z_{\mathbf{k},\mu} \rangle = e^{-\vert z_{\mathbf{k},\mu}\vert^2/2} e^{z_{\mathbf{k},\mu}b^{\dagger}_{\mathbf{k},\mu}}\vert 0 \rangle\,,
\end{align}
where
\begin{align}
\begin{pmatrix} a_{\mathbf{k}} \\ a^{\dagger}_{-\mathbf{k}}\end{pmatrix} =\mathcal{T}^{\mathbf{k}} \begin{pmatrix} b_{\mathbf{k}} \\ b^{\dagger}_{-\mathbf{k}}\end{pmatrix} 
\end{align}
and $b_{\mathbf{k},\mu} \vert z_{\mathbf{k},\mu}\rangle=z_{\mathbf{k},\mu}\vert z_{\mathbf{k},\mu}\rangle$. The average number of magnons in the $\mu$-band with crystal momentum $\mathbf{k}$ is $\bar{n}_{\mathbf{k},\mu}=\langle z_{\mathbf{k},\mu} \vert b^{\dagger}_{\mathbf{k},\mu}b_{\mathbf{k},\mu}\vert z_{\mathbf{k},\mu}\rangle$.
The dynamic correction to the magnetization due to $\bar{n}_{\mu, \mathbf{k}}$ magnons at mode $\mu$ and crystal momentum $\mathbf{k}$ is calculated via magnon coherent states,
    \begin{align}
   \langle \mathbf{S}_{\mathbf{n}, i}\rangle=     \bra{z_{\mathbf{k},\mu}} \mathbf{S}_{\mathbf{n}, i} \ket{z_{\mathbf{k},\mu}}
        = \sqrt{\frac{2S \bar{n}_{\mu, \mathbf{k}}}{N}}\operatorname{Re}\left[\mathbf{v}^-_i \left(\mathcal{T}^{\mathbf{k}}_{i,\mu} e^{i(\mathbf{k} \cdot (\mathbf{R_n}+\mathbf{r_i}) - SE^{\mathbf{k}}_{\mu \mu}t)} + \left(\mathcal{T}^{\mathbf{k}}_{i+N,\mu}\right)^* e^{-i(\mathbf{k} \cdot (\mathbf{R_n}+\mathbf{r_i}) - SE^{\mathbf{k}}_{\mu \mu}t)}\right) \right] + S\mathbf{v}^3_i\,.
    \end{align}

The components of real-space deformation are $\Delta \mathbf{S}_{\mathbf{n}, i}=\bra{z_{\mathbf{k},\mu}} \mathbf{S}_{\mathbf{n}, i} \ket{z_{\mathbf{k},\mu}}-S\mathbf{v}^3_i$.
\begin{figure}[b]
    \centering
    \includegraphics[width=0.9\linewidth]{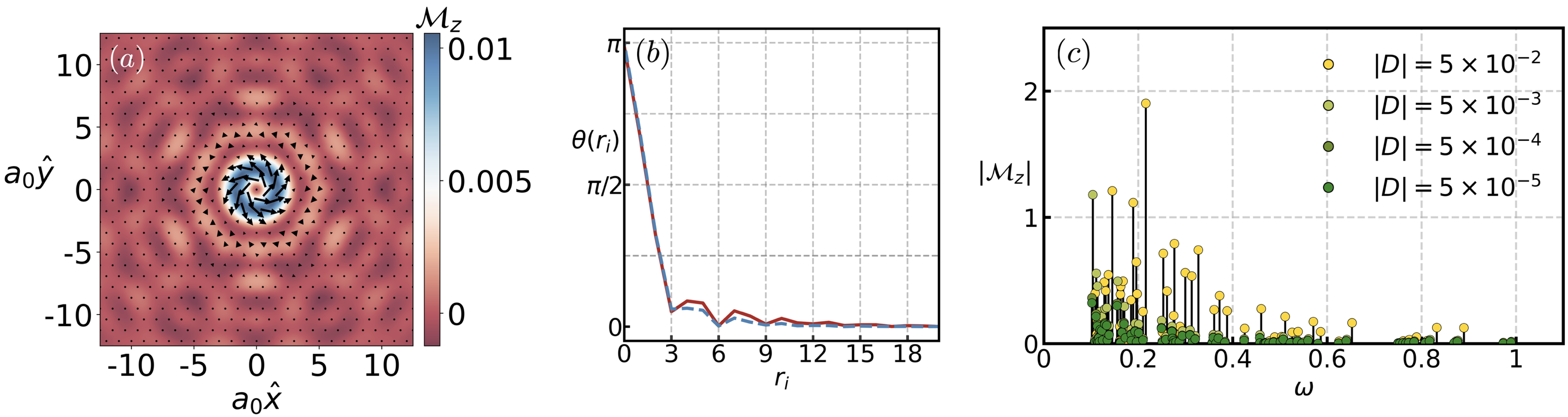}
    \caption{\textbf{Vanishing dipole moment due to skyrmion tail.} (a) Spatial dependence of the magnetic dipole moment $\mathcal{M}_z$ for the lowest breathing mode at $\omega=0.028$. The global spin is conserved, $\mathcal{M}_z=0$, since the out-of-plane dipole moment of the skyrmion core $\sum_{i < \lambda} \mathcal{M}^{i}_z$ cancels out by the opposite dipole moment of the tail, $\sum_{i> \lambda} \mathcal{M}^{i}_z$. (b) Reduction of the skyrmion tail amplitude from the inversion-symmetric ground state (red line) after adding a small DMI of $|\mathbf{D|}=0.05$ (blue line). (c) Non-vanishing $\mathcal{M}_z$ arising due to broken inversion symmetry. The canceling moment from the skyrmion tail is smaller than the contribution from the core. The breathing modes with the largest $\mathcal{M}_z$ are found close to $\omega_{\mbox{\scriptsize gap}}$.}
    \label{dipole_BM}
\end{figure}
\subsection*{C. AC Fields}\label{sec:AC_Fields}

Magnon coherent states can be excited by an applied AC field if they possess a finite magnetic dipole moment calculated by \cite{PRXQuantum.3.040321},
\begin{align}
    \boldsymbol{\mathcal{M}}_\mu = \frac{1}{\sqrt{N}} \sum^{N}_{i} \sqrt{\frac{S}{2}}\left(\mathbf{v}^-_i\mathcal{T}^{\mathbf{k}=0}_{i,\mu} + \mathbf{v}^+_i \mathcal{T}^{\mathbf{k}=0}_{i+N,\mu}\right)\,.
\end{align}
In Fig.~\ref{Isolated_Skyrmion}-(d) and Fig.~\ref{Chern_numbers} we plot $\mathcal{M}_{x,y}$ for an isolated skyrmion and a skyrmion lattice respectively. Modes with a finite $\mathcal{M}_{x,y}$ and frequency $\omega_{\lambda}$ directly couple to in-plane alternating magnetic fields. We perform micromagnetic simulations by numerically solving the LLG equation with an AC magnetic field defined on a discrete lattice using the fourth-order Runge-Kutta method:

\begin{dmath}
    \frac{d\mathbf{S}_i}{dt} = \mathbf{S}_i \times \mathbf{H}^{\mbox{\scriptsize eff}}_i - \alpha \mathbf{S}_i \times (\mathbf{S}_i \times \mathbf{H}^{\mbox{\scriptsize eff}}_i)\,,
\end{dmath}

where $\mathbf{H}^{\mbox{\scriptsize eff}}_i=\frac{\partial H}{\partial \mathbf{S}_i}$ and $\alpha$ is the damping constant. In Figs.~\ref{Localization}-(d),(h), and (l), we show numerical solutions of the dynamical LLG equation under an oscillating magnetic field of the form $\textbf{H}_{\mbox{\scriptsize AC}}=h_{\mbox{\scriptsize AC}} \cos(\omega_{\lambda} t) \hat{\mathbf{x}}$ where we plot deviations of the spin from the initial state, $\mathcal{D}\mathbf{S}_i(t)= \mathbf{S}_{i}(t=0) - \mathbf{S}_i(t)$. 

Note that the emergence of dynamical magnon superlattices is due to the hybridization of the CCW skyrmion modes with finite $\mathcal{M}_{x,y}$ and extended states; simple superpositions of degenerate low-lying extended states displaying the interference patterns of Fig.~\ref{superposition} (a) are not excited under an AC field. Importantly, the superlattice structure persists at large distances from the skyrmion [see Fig.~\ref{superposition} (b)]. As expected, due to the $S_z$-conserving U(1) symmetry of the model, we find $\mathcal{M}_z=0$ for all modes considered. For the lowest breathing mode of the skyrmion at $\omega=0.028$, the localized dipole moment near the skyrmion core is exactly canceled by the delocalized moment of the surrounding ferromagnet (see Figure \ref{dipole_BM} (a)). 

Still, these modes can be excited under an AC field in the $z$-direction through a second-order coupling mechanism beyond the linear response regime. For an isolated skyrmion, the $U(1)$ symmetry of the model implies that the model is invariant under $\mathbf{S}_i \rightarrow R_z(\varphi_0) \mathbf{S}_i$, making skyrmion helicity $\varphi_0$ a zero mode. The conjugate momentum $\mathcal{S}^z=\sum_i(S-S^z_{i})$ is conserved and equal to the skyrmion spin $\mathcal{S}^z_0=\sum_i(S-S^z_{i,0})$ \cite{PhysRevLett.127.067201}. Thus, if we consider small deviations from the skyrmion spin, $S_i^z=S_{i,0}^z+\delta S^z_i$, the constraint of skyrmion spin conservation suggests $\sum_i \delta S^z_i=0$, in agreement with the numerical finding $\mathcal{M}_z=0$ for all modes. An out-of-plane AC field of frequency $\omega$ couples to $\mathcal{S}^z_0$, and induces oscillations of $\varphi_0(t)\sim \cos(\omega t)$. Due to the strong hybridization between breathing modes and helicity [see Fig.\ref{Breathing} (a)-(d)], this time-dependent helicity parametrically modulates the magnon mode profiles, leading to their weak excitation Fig.~\ref{Localization} (p). As a result, the breathing mode dynamics become indirectly activated through mode hybridization. This is a second-order phenomenon beyond linear order, resulting from the backaction of magnons on the background.

\subsection*{D. Rotational Symmetry-Breaking and Chiral Magnets}\label{sec:Chiral_Magnets}

Chiral symmetry-breaking resulting from small perturbative terms (such as the small DMI present in real materials) reintroduces the linear coupling between breathing modes and out-of-plane magnetic fields. Specifically, DMI reduces the amplitude of the skyrmion tail (see Fig.~\ref{dipole_BM} (b)), and its dipole moment no longer cancels that of the core, resulting in finite $\mathcal{M}_z$ for breathing modes (see Fig.~\ref{dipole_BM} (c)). 
\noindent

Superlattices emerge due to the unique dynamics of magnons in the FM state of frustrated magnets and therefore persist after adding DMI and dipolar interactions (see Fig. ~\ref{Symmetry_Break}). For this reason, they cannot be observed around DMI-stabilized skyrmions in chiral magnets (see Fig.~\ref{CCW_DMI}). Dipolar interactions are considered in real space with respect to a monolayer geometry, where the classical and quadratic magnon Hamiltonians are respectively:

 \begin{equation}
    H_{\text{DDI}} = \xi \sum_{i, j} \mathbf{S}_i \cdot \mathbf{G}_{ij} \cdot \mathbf{S}_j\text{, where }\mathbf{G}_{ij}=\frac{3\, \mathbf{r}_{ij} \mathbf{r}_{ij} - r_{ij}^2 \mathbf{I}}{r_{ij}^5}
\end{equation}

\begin{equation}
    H_{\text{DDI, SW}} = \frac{\xi S}{2} \sum_{i, j} \left[\begin{pmatrix} a^\dagger_i & a_i \end{pmatrix} \begin{pmatrix}
        \mathbf{v}^+_i \cdot \mathbf{G}_{ij} \cdot \mathbf{v}^-_j & \mathbf{v}^+_i \cdot \mathbf{G}_{ij} \cdot \mathbf{v}^+_j \\
        \mathbf{v}^-_i \cdot \mathbf{G}_{ij} \cdot \mathbf{v}^-_j & \mathbf{v}^-_i \cdot \mathbf{G}_{ij} \cdot \mathbf{v}^+_j
    \end{pmatrix}
    \begin{pmatrix} a_j \\ a^\dagger_j \end{pmatrix} -\left(\mathbf{v}^3_i \cdot \mathbf{G}_{ij} \cdot \mathbf{v}^3_j \right)\left(a^\dagger_i a_i + a_i a^\dagger_i \right)\right]
\end{equation}

\noindent
The effect of these interactions on superlattices is minimal (see Fig. ~\ref{Symmetry_Break} (a)), though deformation occurs if interactions are strong (see Fig. ~\ref{Symmetry_Break} (b)). Similarly, superlattices do not respond significantly to a DMI term (see Fig. ~\ref{Symmetry_Break} (c)).

 \begin{figure}[b]
     \centering
     \includegraphics[width=1\linewidth]{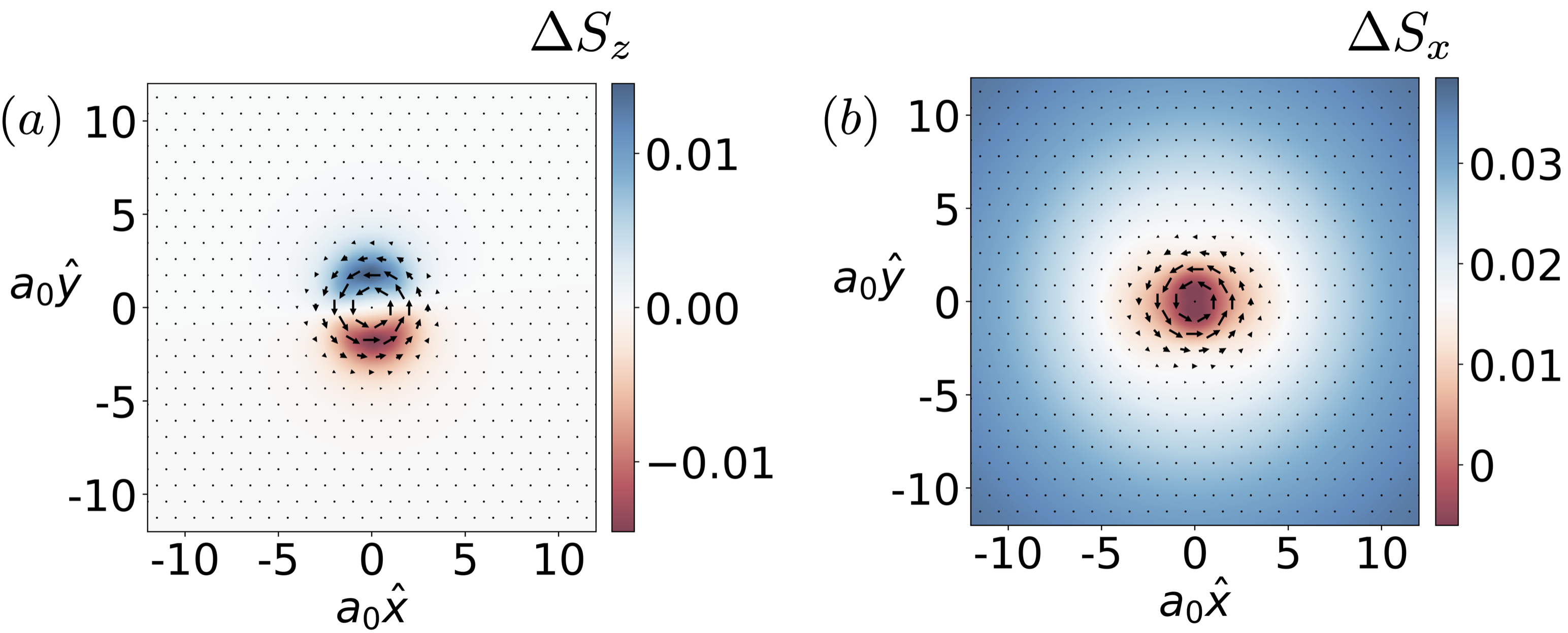}
     \caption{\textbf{Absence of superlattices in chiral magnets.} Real-space deformation (a) $\Delta S_z$ and (b) $\Delta S_x$ of the lowest CCW mode around a skyrmion in a chiral magnet. The crystal-like patterns are completely absent. }
     \label{CCW_DMI}
 \end{figure}

\begin{figure}[b]
    \centering
    \includegraphics[width=1\linewidth]{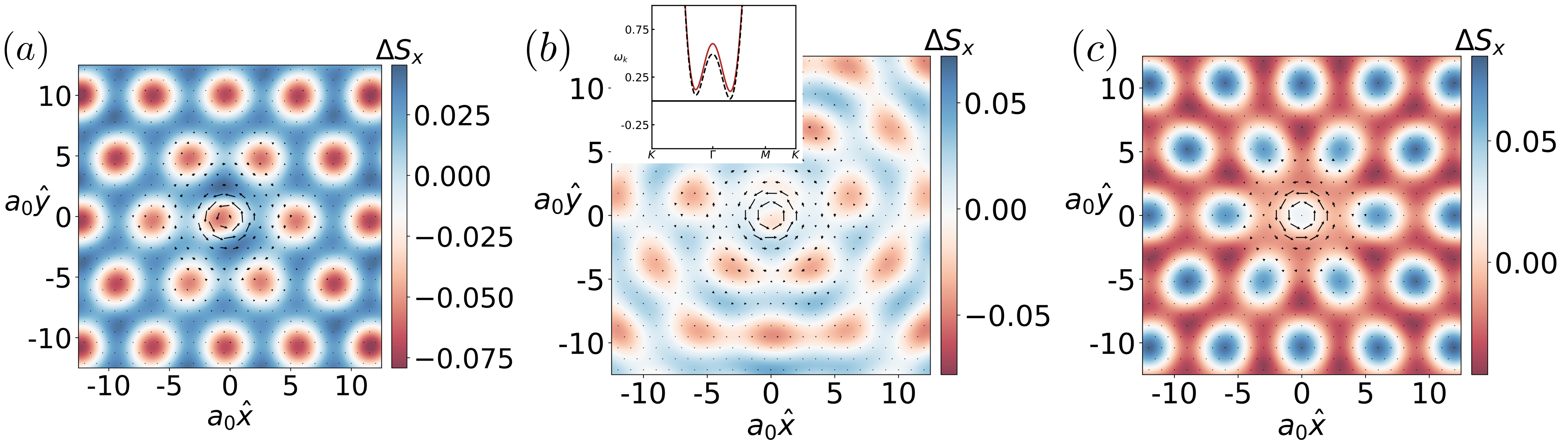}
    \caption{\textbf{Robustness of magnon superlattices under rotational symmetry breaking terms}. Real space deformation $\Delta S_x$ calculated using the model of Eq.~\ref{eq:Hamiltonian} with $J_1=1$, $J_2=0.5$, $K=0.15$ in the presence of rotational symmetry-breaking terms: (a) a weak dipole-dipole term with $\xi=0.001$ and $h=0.225/S$, (b) strong dipole-dipole with $\xi=0.01$ and $h=0.3/S$ and (c) a DMI term of strength $D/J_1=0.01$ and $h=0.225/S$. The inset of (b) depicts the low energy dispersion of magnon excitations around the uniformly magnetized state in the absence (red line) and the presence of strong dipole-dipole interaction $\xi=0.01$ (black line) along the $\Gamma$-M-K direction.}
    \label{Symmetry_Break}
\end{figure}
 
\end{widetext}
\end{document}